\title{A hidden, heavier resonance of the Higgs field}
\author{
        \textsc{Maurizio Consoli} \\ INFN - Sezione di Catania,  I-95129 Catania, Italy \\
        {\tt{maurizio.consoli@ct.infn.it}}\\[0.5cm]
}

\documentclass[12pt]{article}

\usepackage[paper=a4paper,dvips,top=1.5cm,left=1.5cm,right=1.5cm,
    foot=1cm,bottom=1.5cm]{geometry}

\usepackage{times}

\usepackage[fleqn]{amsmath}
\usepackage{amsfonts}
\usepackage{amssymb}
\usepackage{amsthm}
\usepackage{amsopn}
\usepackage{xspace}
\usepackage{array}
\usepackage{epsfig}

\newcommand{\BE}{\begin{equation}}
\newcommand{\EE}{\end{equation}}
\newcommand{\half}{{\scriptstyle{\frac{1}{2}}}}
\begin{document}

\maketitle







%
%
%
%



\begin{abstract}
\noindent In Veltman's original view, the Standard Model with a
large Higgs particle mass $M_h \lesssim$ 1 TeV was the natural
completion of the non-renormalizable Glashow model. In this sense,
this mass was a second threshold for weak interactions, as the W
mass was for the non-renormalizable 4-fermion V-A theory. Today,
after the observation of the narrow scalar resonance with $m_h=$ 125
GeV, Veltman's large $M_h$ seems to be ruled out. Yet, depending on
the description of SSB in $\Phi^4$ theory, this is not necessarily
true. In fact, besides the mass $m_h$ describing its quadratic
shape, the effective potential might exhibit a much larger mass
scale $M_h$ associated with the zero-point energy which determines
its depth. This larger $M_h$ controls vacuum stability and,
differently from $m_h$, would remain finite in units of the weak
scale $\langle \Phi \rangle\sim$ 246.2 GeV for infinite ultraviolet
cutoff. Lattice simulations of the propagator are consistent with
this two-mass structure and lead to the estimate $M_h\sim 700$ GeV.
In spite of its large mass, however, the heavier state would couple
to longitudinal W's with the same typical strength of the low-mass
state and thus represent a relatively narrow resonance. In this way,
such hypothetical resonance would naturally fit with some excess of
4-lepton events observed by ATLAS around 680 GeV. Analogous data
from CMS are needed to confirm or disprove this interpretation.
Implications of this two-mass structure for radiative corrections
will also be discussed.
\end{abstract}
\vfill\eject

\section{Introduction}

In principle, there were many possible scenarios for the Higgs
particle mass. At the extremes of the mass range one could consider
two basically different options. The (minimal) supersymmetry, where
the mass of the lightest Higgs scalar is $m_h \sim M_w$, and
Veltman's idea \cite{veltman1977} of a large $M_h \lesssim 1$ TeV.
In his view, incorporating the idea of Spontaneous Symmetry Breaking
(SSB) \cite{Englert:1964et,Higgs:1964ia} with a large $M_h$, the
Standard Model was the natural completion of the non-renormalizable
Glashow model \cite{glashow1961}. In this sense, he was speaking of
a second threshold in weak interactions, just like with the W mass
and the non-renormalizable 4-fermion V-A theory. Today, after the
observation at LHC \cite{Aad:2012tfa,Chatrchyan:2012xdj} of the
narrow scalar resonance with mass $m_h \sim 125 $ GeV, Veltman's
large $M_h$ seems to be definitely ruled out.

However, this is not necessarily true. So far, only the gauge and
Yukawa couplings of the 125 GeV resonance have been tested. The
effects of a genuine scalar self-coupling $\lambda= 3 m^2_h/\langle
\Phi \rangle^2$ are still below the accuracy of the measurements and
an uncertainty about the origin of SSB still persists. Given this
uncertainty, one may wonder about the traditional assumption that
the Higgs field propagator has only one pole associated with the
quadratic shape of the scalar potential at its minima. While
Goldstone bosons are well understood, depending on the physical
mechanisms which induce SSB, the broken-symmetry phase could display
a richer pattern of scales and the Higgs field exhibit some yet
undiscovered, heavier resonance of mass $M_h \gg m_h$. Veltman's
original view could then be realized albeit in a new, unexpected
way.

About the origin of SSB, at the beginning the driving mechanism was
just a classical potential with double-well shape. Later on, after
Coleman and Weinberg \cite{Coleman:1973jx}, this classical potential
was replaced by the effective potential $V_{\rm eff}(\varphi )$
which, in principle, includes the zero-point energy of all particles
in the spectrum. But SSB could still be determined by the pure
scalar sector if the other contributions to the vacuum energy were
negligible. Then, as argued in refs.\cite{Consoli:2020nwb,symmetry},
a series of logical steps leads to the idea of a Higgs propagator
with more than one peak.

One should first follow those lattice simulations of $\Phi^4$ in 4D
\cite{lundow2009critical,Lundow:2010en,akiyama2019phase} indicating
that SSB is a (weak) first-order phase transition. While in the
presence of gauge bosons SSB is often described as a first-order
transition, in pure $\Phi^4$ this requires to replace standard
perturbation theory with some alternative scheme. A scheme where
massless $\Phi^4$ (i.e. classically scale invariant) exhibits SSB so
that the phase transition occurs earlier, when the quanta of the
symmetric phase have a tiny but still positive mass squared. One
then discovers that, in the two easily available alternative schemes
(the simple 1-loop and/or Gaussian approximation), $V_{\rm
eff}(\varphi )$ has {\it two} distinct mass scales
\cite{Consoli:2020nwb,symmetry}:~i) a mass $m^2_h$, defined by its
quadratic shape at the minimum and~ ii) a mass $M^2_h$ entering the
zero-point energy which determines its depth. Always considered as
being the same mass, in these approximations one finds instead
$M^2_h\sim L m^2_h \gg m^2_h$, where $L=\ln (\Lambda_s/M_h)$ and
$\Lambda_s$ is the ultraviolet cutoff of the scalar sector. Since
vacuum stability depends on the much larger $M_h$, and not on $m_h$,
SSB could originate within the pure scalar sector regardless of the
other parameters of the theory (e.g. the vector boson and top quark
mass).

It is obvious that quadratic shape and depth of the potential are
different quantities. At a more formal level, one should recall that
the derivatives of the effective potential produce (minus) the
n-point functions at zero external momentum. Hence $m^2_h$, which is
$V''_{\rm eff}(\varphi )$ at the minimum, is directly the 2-point,
self-energy function $|\Pi(p=0)|$. On the other hand, the zero-point
energy is (one-half of) the trace of the logarithm of the inverse
propagator $G^{-1}(p)=(p^2-\Pi(p))$. Therefore, after subtracting
constant terms and quadratic divergences, matching the 1-loop
zero-point energy(``$zpe$'') gives the relation \BE\label{general}
zpe\sim -\frac{1}{4} \int^{p_{\rm max}}_{p_{\rm min}} {{ d^4
p}\over{(2\pi)^4}} \frac{\Pi^2(p)}{p^4} \sim-\frac{ \langle
\Pi^2(p)\rangle }{64\pi^2} \ln\frac{p^2_{\rm max}}{p^2_{\rm
min}}\sim -\frac{M^4_h}{64\pi^2} \ln\frac{\Lambda^2_s }{M^2_h} \EE
This shows that $M^2_h$ effectively includes the contribution of all
momenta and reflects a typical average value $|\langle
\Pi(p)\rangle| $ at larger $p^2$. A non-trivial momentum dependence
of $\Pi(p)$ would then indicate the coexistence in the
broken-symmetry phase of two kinds of ``quasi-particles'', with
masses $m_h$ and $M_h$, and thus closely resemble the two branches
(phonons and rotons) in the energy spectrum of superfluid He-4 which
is usually considered the non-relativistic analog of the broken
phase.

Actually, due to their conceptual difference, the issue $m_h \neq
M_h$ could be posed on a pure hypothetical basis, independent of any
specific calculation. Then, a non uniform scaling of the two masses
with $\Lambda_s$ could be deduced for consistency with the
``triviality'' of $\Phi^4$. This implies a continuum limit with a
Gaussian set of Green's functions and with just a massive free-field
propagator. Therefore, with this constraint, besides the usual
$m_h\to M_h$ limit, a consistent cutoff theory can also predict
their non-uniform scaling when $\Lambda_s \to \infty$. The type of
single-mass limit will then depend on the unit scale, $m_h$ or
$M_h$, chosen for measuring momenta. Namely: a) $m_h$ is the unit
scale so that $M_h$ and the higher branch simply decouple b) $M_h$
is the unit scale so that, when $\Lambda_s \to \infty$, the phase
space of the lower branch becomes smaller and smaller until ideally
shrinking to the zero-measure set $p_\mu=0$ which is transformed
into itself under the Lorentz Group. This means that the lower
branch merges into the vacuum state and the only remaining
excitation is the higher branch with mass $M_h$, reproducing the
usual picture of a massive fluctuation field for any $p_\mu \neq 0$.

Given this consistency with rigorous field-theoretical results, the
existence of a two-mass structure in the cutoff theory was checked
with lattice simulations of the scalar propagator
\cite{Consoli:2020nwb} in the 4D Ising limit of the theory. This
corresponds to a $\Phi^4$ with an infinite bare coupling
$\lambda_0=+\infty$, as when sitting precisely at the Landau pole.
For a given non-zero, low-energy coupling $\lambda \sim 1/L$, this
represents the best possible definition of the local limit with a
cutoff. Then, once $m^2_h$ is directly computed from the $p^2\to 0$
limit of $G(p)$ and $M^2_h$ is extracted from its behaviour at
higher $p^2$, the lattice data are consistent with a transition
between two different regimes and with the expected increasing
logarithmic trend $M^2_h\sim L m^2_h$.

If, for finite $\Lambda_s$, the scalar propagator really
interpolates between two vastly different scales, by increasing the
energy there should be a transition from a relatively low value,
e.g. $m_h$=125 GeV, to a much larger $M_h$. At the same time,
differently from $m_h$, the larger mass $M_h$ would remain finite in
units of the weak scale $\langle \Phi \rangle\sim (G_{\rm
Fermi}\sqrt{2})^{-1/2}\sim$ 246.2 GeV in the continuum limit. In
fact, by expressing the proportionality relation in terms of some
constant $c_2$, say \BE \label{c2} M^2_h = m^2_h L \cdot(c_2)^{-1}
\EE and replacing the leading-order estimate $\lambda\sim
16\pi^2/(3L)$ in the relation $\lambda= 3 m^2_h/\langle \Phi
\rangle^2$, one obtains a proportionality relation through a
constant $K$ \BE M_h=K \langle \Phi \rangle\EE with $K \sim
(4\pi/3)\cdot (c_2)^{-1/2}$. Since, from a fit to the lattice
propagator \cite{Consoli:2020nwb}, we found $(c_2)^{-1/2} = 0.67 \pm
0.01 ~({\rm stat}) \pm 0.02 ~({\rm sys})$ this gives the
leading-order estimate $M_h = 690 \pm 10 ~({\rm stat}) \pm 20 ~({\rm
sys})~ {\rm GeV}$. Instead, with the next-to-leading $m_h-\langle
\Phi \rangle$ relation and the same $c_2$, we obtained $M_h = 750\pm
10 ~({\rm stat}) \pm 20 ~({\rm sys})$ GeV \cite{Consoli:2020nwb}.
The two values could then be summarized into a final estimate $M_h
\sim 720 \pm 30 $ GeV which accounts for this theoretical
uncertainty and updates the previous work of refs.\cite{Cea:2003gp,
Cea:2009zc}.

I emphasize that, by accepting the ``triviality'' of the theory in
4D, the cutoff-independent combination $3M^2_h/ \langle \Phi
\rangle^2= 3K^2$ cannot represent a measure of observable
interactions. This $3K^2=O(10)$, which is clearly quite distinct
from the other coupling $\lambda=3 m^2_h/ \langle \Phi \rangle^2\sim
1/L$, should not be viewed as a coupling constant which produces
{\it observable} interactions in the broken-symmetry phase. Instead,
since $M_h$ reflects the magnitude of the vacuum energy density, it
would be natural to consider $3K^2\sim \lambda L$ as a {\it
collective} self-interaction of the vacuum condensate which persists
in the $\Lambda_s\to \infty$ limit. This original view
\cite{Consoli:1993jr,Consoli:1997ra} can intuitively be formulated
in terms of a scalar condensate whose increasing density $\sim L$
\cite{Consoli:1999ni} compensates for the decreasing strength
$\lambda \sim 1/L$ of the two-body coupling. On the other hand
$\lambda\sim 1/L$ remains as the appropriate coupling to describe
the {\it individual} interactions of the elementary excitations of
the vacuum, i.e. the Higgs field and the Goldstone bosons. In this
way, consistently with the ``triviality'' of $\Phi^4$ theory, their
interactions will become weaker and weaker for $\Lambda_s \to
\infty$. On this basis, the heavier state would couple to
longitudinal vector bosons with the same typical strength $\lambda=3
m^2_h/ \langle \Phi \rangle^2\sim 1/L$ of the low-mass state and
would thus represent a relatively narrow resonance.

In the following, I will first resume in Sects. 2, 3 and 4 the main
analytical and numerical arguments in favor of the $m_h-M_h$
picture. Later on, in Sects.5 and 6, I will concentrate on more
phenomenological aspects and argue that the hypothetical, heavier
$M_h$ would naturally fit with some excess of 4-lepton events
observed by ATLAS around 680 GeV. Finally in the conclusive Sect.7,
after a short summary, I will mention a work of van der Bij
\cite{jochum2018} where a Higgs field propagator with more than one
peak was also considered. This brings in touch with the possible
effects of a two-mass structure in radiative corrections.

\section{SSB and the effective potential}

To start with, let us recall the description of SSB as a
second-order transition and follow the Particle Data Group (PDG)
\cite{Tanabashi:2018oca} where the scalar potential is expressed as
\BE \label{VPDG} V_{\rm PDG}(\varphi)=-\frac{ 1}{2} m^2_{\rm PDG}
\varphi^2 + \frac{ 1}{4}\lambda_{\rm PDG}\varphi^4 \EE By fixing
$m_{\rm PDG}\sim$ 88.8 GeV and $\lambda_{\rm PDG}\sim 0.13$, this
has a minimum at $|\varphi|=\langle \Phi \rangle\sim$ 246 GeV and
quadratic shape $V''_{\rm PDG}(\langle \Phi \rangle)=$ (125
GeV)$^2$. As a built-in relation, the second derivative of the
potential also determines its depth, i.e. the vacuum energy ${\cal
E}_{\rm PDG}$
 \BE \label{EPDG} {\cal E}_{\rm PDG}=-\frac{ 1}{2} m^2_{\rm PDG}
\langle \Phi\rangle^2 + \frac{ 1}{4}\lambda_{\rm PDG} \langle
\Phi\rangle^4 =-\frac{ 1}{8} (125~{\rm GeV}\langle \Phi\rangle)^2
\sim -1.2\cdot10^8~ {\rm GeV}^4\EE But, as anticipated in the
Introduction, recent lattice simulations of $\Phi^4$ in 4D
\cite{lundow2009critical,Lundow:2010en,akiyama2019phase} indicate
instead a (weakly) first-order transition. SSB would then emerge as
a true instability of the symmetric vacuum at $\varphi=0$. Its
quanta have a tiny and still positive mass squared $V''_{\rm
eff}(\varphi=0)=m^2_\Phi>0$ but nevertheless, below a critical value
$0 \leq m^2_\Phi <m^2_c$, their attractive, long-range interaction
\cite{Consoli:1999ni} can destabilize this symmetric vacuum. The
lowest energy state of the massless theory at $m^2_\Phi=0$ would
then correspond to the broken-symmetry phase, as suggested by
Coleman and Weinberg \cite{Coleman:1973jx} in their original 1-loop
calculation.

In this interpretation of SSB, the dynamics of the symmetric phase
represents the primary $\Phi^4$ sector and its degree of locality
$\Lambda_s$ is the ultimate cutoff scale of the theory (as with the
hard-sphere core of He-4 atoms in superfluid helium). We are thus
lead to identify $\Lambda_s$ as the Landau pole where the bare
coupling $\lambda_0\to +\infty$. This corresponds precisely to the
Ising limit and provides the best possible definition of a local
$\Phi^4$ theory for any non-zero low-energy coupling $\lambda\sim 1/
\ln \Lambda_s\ll 1$ \footnote{This is true at 1-loop. Beyond 1-loop,
standard perturbation theory gives contradictory indications (Landau
pole in odd orders vs. spurious ultraviolet fixed points in even
orders). Borel re-summation procedures \cite{shirkov,chetyrkin,
kazakov}, yielding a positive, monotonically increasing
$\beta-$function, support again the idea of the Landau pole.}. This
latter coupling is instead appropriate for low-energy physics, as in
the original Coleman-Weinberg calculation of the effective potential
for field values $|\varphi| \ll \Lambda_s$ \footnote{By assuming a
Landau pole, the $\lambda$ in the effective potential is naturally
interpreted as the small coupling at a scale $\mu \sim |\varphi|\ll
\Lambda_s$. However, rejecting the Landau pole,from the resulting
trend $\lambda\sim 1/ \ln \Lambda_s$, this $\lambda$ could also be
interpreted as an infinitesimal ``asymptotically free'' bare
coupling. In the more general context of the $\epsilon-$expansion,
the two points of view might reflect the existence of two separate
$\Phi^4$ theories living in $D = 4 \pm \epsilon$ space-time
dimensions \cite{stevenson1987,mpla1996}. }. The Coleman-Weinberg
calculation represents the simplest scheme which is consistent with
a weak first-order picture. We will first reproduce below this well
known computation and discuss afterward its general validity. After
subtraction of constant terms and quadratic divergences, the
effective potential is \BE V_{\rm eff}(\varphi) = \frac{\lambda}{4!}
\varphi^4 +\frac{\lambda^2}{256 \pi^2} \varphi^4 \left[ \ln (\half
\lambda \varphi^2 /\Lambda^2_s ) - \frac{1}{2} \right]  \EE and its
first few derivatives are \BE \label{vprime}
V'_{\rm eff}(\varphi) =  \frac{\lambda}{6} \varphi^3
+\frac{\lambda^2}{64 \pi^2} \varphi^3 \ln (\half \lambda \varphi^2
/\Lambda^2_s )  \EE and \BE \label{vsecond}
V''_{\rm eff}(\varphi) =  \frac{\lambda}{2} \varphi^2
+\frac{3\lambda^2}{64 \pi^2} \varphi^2 \ln (\half \lambda \varphi^2
/\Lambda^2_s ) +\frac{\lambda^2\varphi^2}{32\pi^2}   \EE By
introducing the mass squared parameter $M^2(\varphi)\equiv \half
\lambda\varphi^2$, the same potential can be expressed as a
classical background + zero-point energy of a particle with mass
$M(\varphi)$, i.e. \BE \label{zero} V_{\rm eff}(\varphi) =
\frac{\lambda\varphi^4}{4!} - \frac{M^4(\varphi) }{64 \pi^2}  \ln
\frac{ \Lambda^2_s \sqrt{e} } {M^2(\varphi)}   \EE Thus, non-trivial
minima of $V_{\rm eff}(\varphi)$ occur at those points $\varphi=\pm
v$ where \BE \label{basic} M^2_h={{\lambda v^2}\over{2
}}=\Lambda^2_s \exp( -{{32 \pi^2 }\over{3\lambda }})\EE with a
quadratic shape \BE \label{mh} m^2_h\equiv V''_{\rm eff}(\pm v) =
\frac{\lambda^2v^2}{32\pi^2}= \frac{\lambda}{16\pi^2}M^2_h\sim
\frac{M^2_h}{L} \ll M^2_h \EE where $L\equiv \ln
\frac{\Lambda_s}{M_h}$.  Notice that the energy density depends on
$M_h$ and {\it not} on $m_h$, because \BE \label{basicground} {\cal
E}= V_{\rm eff}(\pm v)= -\frac{M^4_h}{128 \pi^2 } \EE therefore the
critical temperature at which symmetry is restored, $k_BT_c\sim
M_h$, and the stability of the broken phase depends on the larger
$M_h$ and not on the smaller $m_h$.

Now, one may object to the above straightforward minimization
procedure that the 1-loop calculation is just the first term of an
infinite series and should be further ``improved''. As it is well
known, in this conventional view, the 1-loop minimum disappears and
one would again predict a second-order transition, just the result
that we know to be in contrast with most recent lattice simulations.
Therefore, one should look at the calculation in the different
perspective of Eq.(\ref{zero}). This has the qualitatively different
meaning of a classical background + zero-point energy with a
$\varphi-$dependent mass and, as such, is consistent by itself
without any need of being further improved.

To confirm the validity of this interpretation, one can compare with
other approximation schemes, for instance the Gaussian approximation
\cite{Barnes:1978cd,Stevenson:1985zy} which has a variational nature
and explores the Hamiltonian in the class of the Gaussian functional
states. It also represents a very natural alternative because, at
least in the continuum limit, a Gaussian structure of Green's
functions fits with the generally accepted ``triviality'' of the
theory in 4D. This other calculation produces a result in agreement
with the one-loop potential \cite{Consoli:1993jr,Consoli:1997ra}.
This is not because there are no non-vanishing corrections beyond
1-loop; there is actually an infinite resummation of terms. The
point, however, is that those additional terms do not alter the
functional form of the result which is the same as in
Eq.(\ref{zero}) \BE \label{vgauss}
V^G_{\rm eff}(\varphi) =  \frac{\hat\lambda\varphi^4}{4!}
-\frac{\Omega^4(\varphi) }{64 \pi^2}  \ln \frac{ \Lambda^2_s
\sqrt{e} } {\Omega^2(\varphi)} \EE with \BE \hat \lambda=
\frac{\lambda } {1 + \frac{\lambda}{16 \pi^2} \ln \frac {\Lambda_s}{
\Omega(\varphi)}   }  ~~~~~~~~~~~~ {\rm and}~~~~~~~~~~~~~~~~~
\Omega^2(\varphi) = \frac{\hat\lambda\varphi ^2} {2 } \EE This
explains why the one-loop potential can also admit a
non-perturbative interpretation. It is the prototype of the Gaussian
and of the {\it infinite number} of ``post-gaussian'' calculations
\cite{Stancu:1989sk,Cea:1996pe} where higher-order contributions are
effectively reabsorbed into the same basic structure: classical
background + zero-point energy with a $\varphi-$dependent mass.

\section{Eliminating $\Lambda_s$ in the $M_h-\langle \Phi \rangle$ relation}

The effective potential of Sect.2 provides a different path to
renormalization. Since, for any non-zero $\lambda$, there is a
finite Landau pole, one could in fact consider the whole set of
pairs ($\Lambda_s$,$\lambda$),($\Lambda'_s$,$\lambda'$),
($\Lambda''_s$,$\lambda''$)...with different Landau poles and
corresponding low-energy couplings. By considering this whole set of
parameters, and imposing some symmetry principle, one can minimize
the influence of the cutoff on observable quantities and even
consider the $\Lambda_s\to \infty$ limit.

The basic constraint on the equivalent ($\Lambda_s$,$\lambda$)
pairs, consists in requiring the same vacuum energy
Eq.(\ref{basicground}), or equivalently the same mass scale
Eq.(\ref{basic}), namely \BE \label{CSground}
\left(\Lambda_s\frac{\partial}{\partial\Lambda_s} +
\Lambda_s\frac{\partial \lambda}{\partial\Lambda_s}\frac{\partial
}{\partial \lambda}\right){\cal E}(\lambda,\Lambda_s)=0 \EE With the
definition \BE \Lambda_s\frac{\partial
\lambda}{\partial\Lambda_s}\equiv -\beta(\lambda)=
-\frac{3\lambda^2}{16\pi^2} +O(\lambda^3)  \EE this gives $|{\cal
E}| \sim {\cal I }^4_1$, where ${\cal I }_1$ is the first
RG-invariant \footnote{ Note the minus sign in the definition of the
$\beta-$ function. This is because, in our coupling constant
$\lambda\equiv\lambda(\mu,\Lambda_s)$, at $\mu \sim \varphi$, we are
differentiating with respect to the cutoff and not with respect to
$\mu$. Thus, at fixed $\mu$, $\lambda$ has to decrease by increasing
$\Lambda_s$.} \BE \label{I1} {\cal I }_1= M_h= \Lambda_s
\exp({\int^{\lambda}\frac{dx}{\beta(x)} })\sim \Lambda_s \exp( -{{16
\pi^2 }\over{3\lambda }})\EE The above relations derive from the
more general invariance of the effective potential in the
three-dimensional space ($\varphi$, $\lambda$, $\Lambda_s$) \BE
\label{CSveff} \left(\Lambda_s\frac{\partial}{\partial\Lambda_s} +
\Lambda_s\frac{\partial \lambda}{\partial\Lambda_s}\frac{\partial
}{\partial \lambda}  + \Lambda_s\frac{\partial
\varphi}{\partial\Lambda_s}\frac{\partial }{\partial \varphi}
\right)  V_{\rm eff}(\varphi,\lambda,\Lambda_s)=0 \EE In fact, at
the minima $\varphi=\pm v$, where $(\partial V_{\rm eff}/\partial
\varphi)= 0$, Eq.(\ref{CSground}) is a direct consequence of
Eq.(\ref{CSveff}). Another consequence of this analysis is that,
besides a first invariant mass scale ${ \cal I}_1= M_h$, by
introducing an anomalous dimension for the vacuum field \BE
\label{anomalous} \Lambda_s\frac{\partial
\varphi}{\partial\Lambda_s}\equiv \gamma(\lambda) \varphi \EE there
is a second invariant, namely \BE \label{I2} {\cal I }_2(\varphi) =
\varphi \exp({\int^{\lambda}dx \frac{\gamma(x)}{\beta(x)} })\EE
which introduces a particular normalization of $\varphi$. This had
to be expected because from Eq.(\ref{basic}) the cutoff-independent
combination is $\lambda v^2\sim M^2_h={\cal I }^2_1$ and not $v^2$
itself implying $\gamma= \beta/(2\lambda)$. This particular
definition of the average vacuum field \footnote{ This somewhat
resembles the definition of the physical gluon condensate in QCD
which is $\langle g^2 F^a_{\mu\nu} F^{a\mu\nu} \rangle$ and not just
$\langle F^a_{\mu\nu} F^{a\mu\nu} \rangle$.} is then the natural
candidate to represent the weak scale \cite{Consoli:2020nwb} \BE {
\cal I}_2(v)=\langle \Phi \rangle\sim 246.2 ~ {\rm GeV} \EE so that
the minimization of the effective potential can be expressed as a
proportionality relation of the two invariants ${ \cal I}_1$ and ${
\cal I}_2$ through some constant $K$, i.e. \BE M_h = K \langle \Phi
\rangle\EE On the other hand, the second derivative at the minima,
$m^2_h \equiv (\partial^2 V_{\rm eff}/\partial \varphi^2)$ at
$\varphi=\pm v$, remains as a cutoff-dependent quantity.

With such guiding principle from the effective potential, one
deduces that $M_h$ and $\langle \Phi \rangle$ scale uniformly with
$\Lambda_s$. The constant $K$ could then be extracted from a lattice
simulation of the propagator, by combining the $M_h/m_h$ ratio with
a theoretical $m_h-\langle \Phi \rangle$ relation. The main
ingredients of this analysis will be reported in Sect.4.

\section{Lattice simulation of the propagator}

To show that the existence of two mass scales in the broken phase is
not just speculation, let us now compare with lattice simulations of
the propagator. These were performed \cite{Consoli:2020nwb} in the
4D Ising limit of the theory which has always been considered a
convenient laboratory to exploit the non-perturbative aspects of the
theory. As anticipated, it corresponds to a $\Phi^4$ with an
infinite bare coupling $\lambda_0=+\infty$, as if one were sitting
precisely at the Landau pole. In this sense, for any finite cutoff
$\Lambda_s$, it provides the best definition of the local limit for
a given non-zero, low-energy coupling $\lambda \sim 1/L$ (where
$L=\ln (\Lambda_s/M_h)$).

\begin{figure}[ht]
\centering
\includegraphics[width=0.50\textwidth,clip]{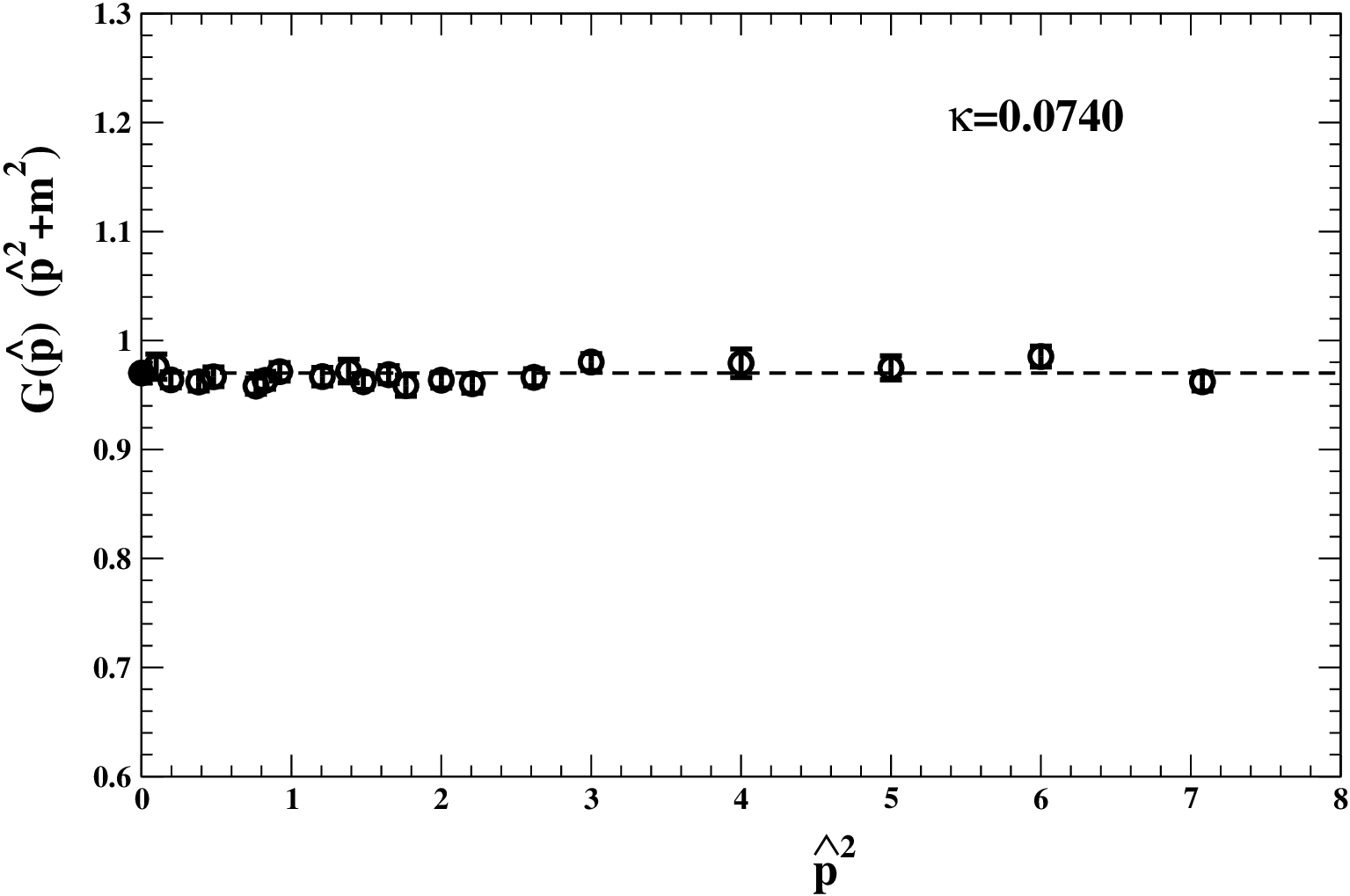}
\caption{\it The lattice data of ref.\cite{Consoli:2020nwb} for the
re-scaled propagator in the symmetric phase at $\kappa=0.074$ as a
function of the square lattice momentum ${\hat p}^2$. The fitted
mass from high ${\hat p}^2$, $m_{\rm{latt}}=0.2141(28)$, describes
well the data down to ${\hat p}=0$. The dashed line indicates the
value of $Z_{\rm{prop}}=0.9682(23)$ and the ${\hat p}=0$ point is
$2\kappa\chi m^2_{\rm{latt}}=0.9702(91)$. } \label{0.074}
\end{figure}

Let us start from the traditional Euclidean lattice action of
$\Phi^4$ theory \BE S= \sum_x \left[ \frac{1}{2}\sum^4_{\mu=1}
(\partial_\mu \Phi_0)^2 + \frac{1}{2} m^2_0 \Phi^2_0(x) +
\frac{\lambda_0}{4!}\Phi^4_0(x)\right] \EE where $\partial_\mu
\Phi_0= \Phi_0( x +\hat \mu) - \Phi_0( x)$ and the lattice spacing
is taken $a=1$. Analogously, all masses are given in units of $1/a$
and the cutoff is $\Lambda_s=\pi/a$. In this action, let us perform
the following changes of variables \cite{Stevenson:2005yn} $\Phi_0=
\sqrt{2\kappa} \Phi$, $m^2_0= (1-2g)/\kappa -8$ and $\lambda_0=
6g/\kappa^2$ so that we obtain \BE S= \sum_x \left[ -2\kappa
\sum^4_{\mu=1}\Phi( x +\hat \mu)\Phi(x) + \Phi^2(x) + g (\Phi^2(x)
-1)^2 \right] \EE Then, for $\lambda_0 \to +\infty$, the lattice
field can only take the values $\Phi(x)=\pm 1$ and one gets the
Ising limit \BE S_{Ising}= -2\kappa \sum_x \sum^4_{\mu=1}\Phi( x
+\hat \mu)\Phi(x)\EE the broken-symmetry phase corresponding to
$\kappa> \kappa_c$, with $\kappa_c=0.0748474(3)$
\cite{lundow2009critical,Lundow:2010en}. With this lattice action,
we computed the lattice field vacuum expectation value
\begin{equation}
\label{baremagn}
 v=\langle |\Phi| \rangle \quad , \quad \Phi \equiv \frac{1}{V_4}\sum_x
\Phi(x)
\end{equation}
and the connected propagator
\begin{equation}
\label{connected} G(x)= \langle \Phi(x)\Phi(0)\rangle - v^2
\end{equation}
where $\langle ...\rangle$ denotes averaging over the lattice
configurations.

\begin{figure}[ht]
  \centering
  \includegraphics[width=0.50\textwidth,clip]{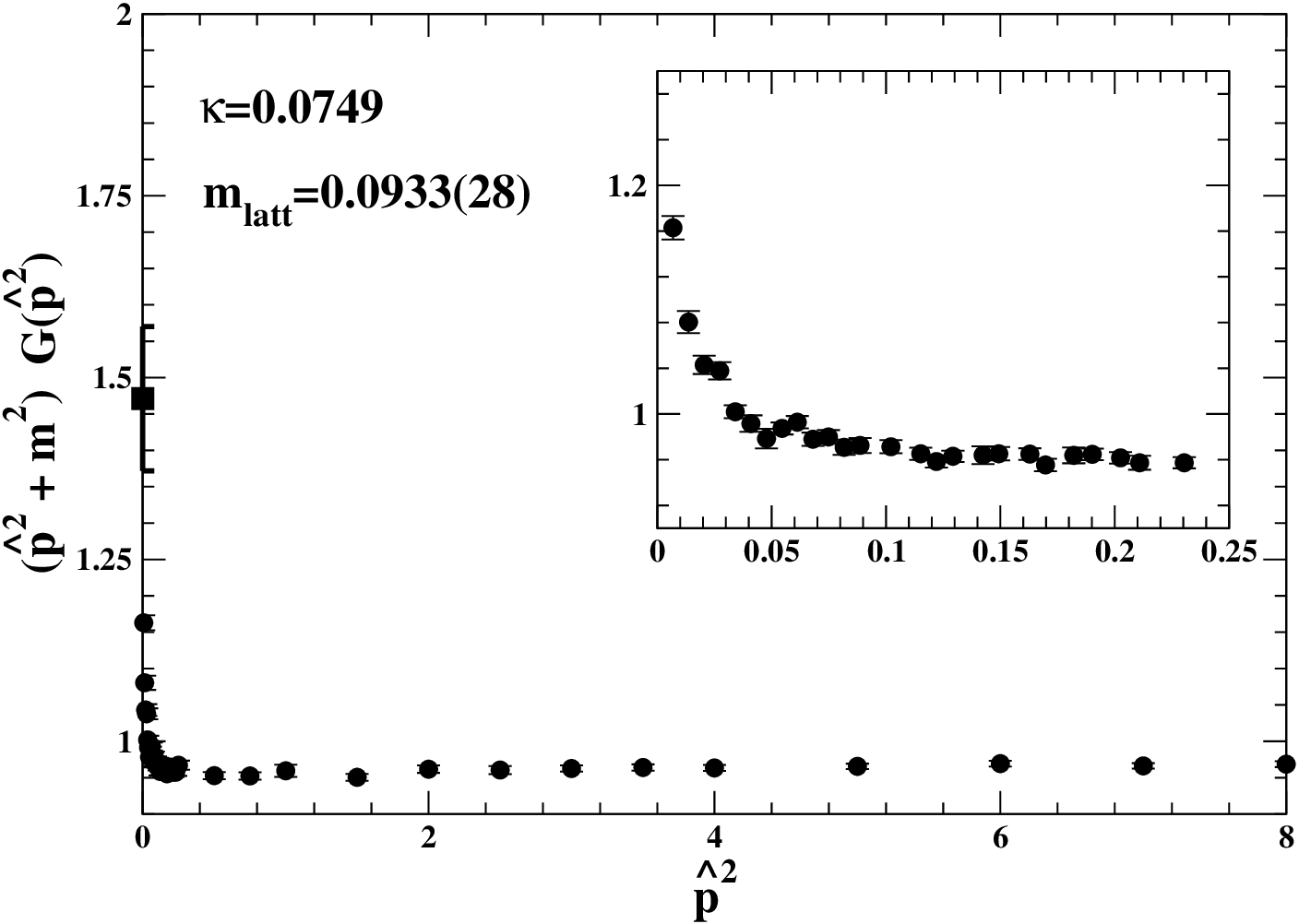}
  \caption{\it The propagator data of ref.\cite{Consoli:2020nwb}, for
$\kappa=0.0749$, rescaled with the lattice mass $M_h \equiv
m_{\rm{latt}}=0.0933(28)$ obtained from the fit to all data with
${\hat p}^2>0.1$. The peak at $p=0$ is $M^2_h/m^2_h= 1.47(9)$ as
computed from the fitted $M_h$ and
$m_h=(2\kappa\chi)^{-1/2}=0.0769(8)$.} \label{0933}
\end{figure}

\begin{figure}[ht]
  \centering
   \includegraphics[width=0.50\textwidth,clip]{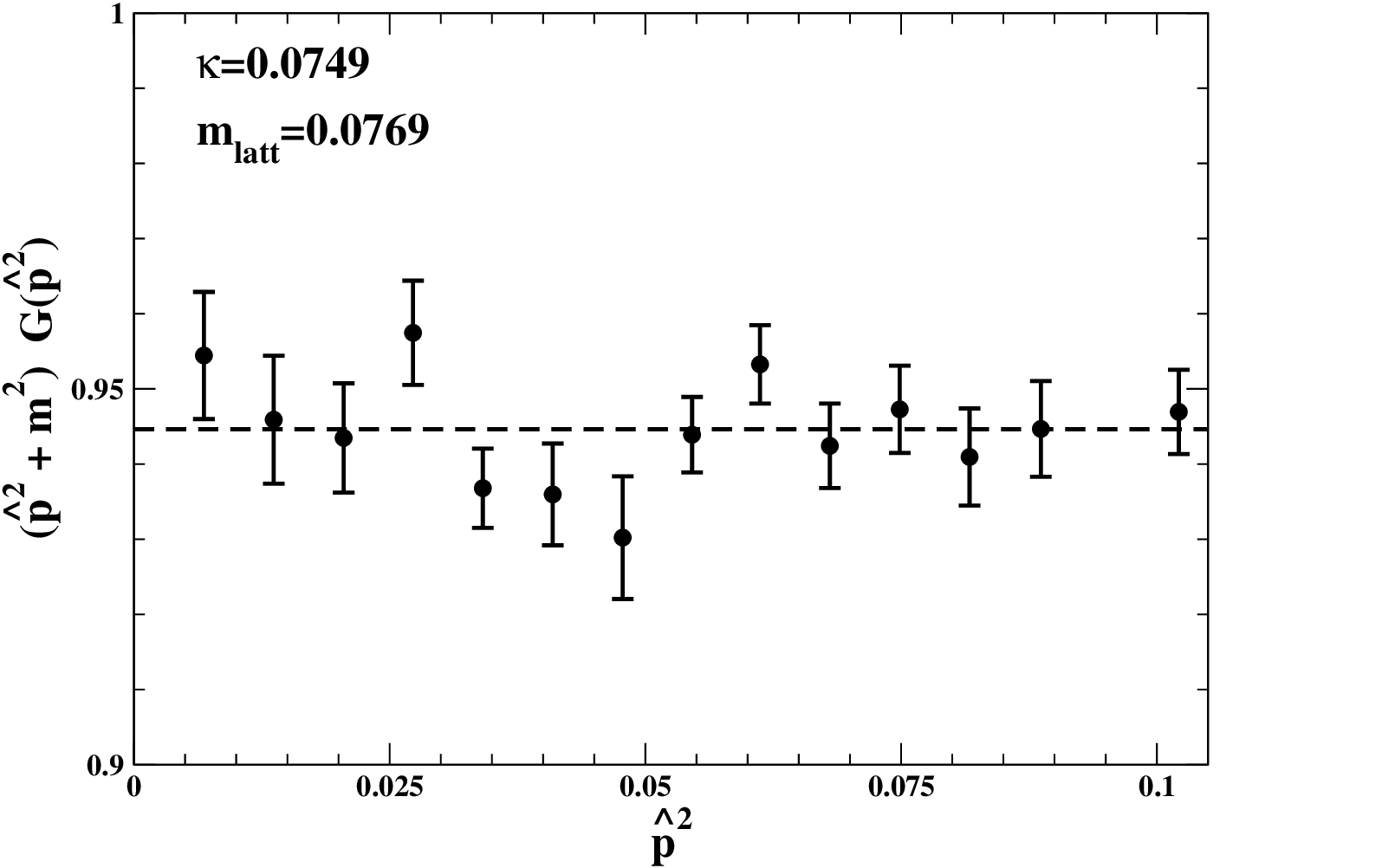}
   \caption{\it The propagator data of ref.\cite{Consoli:2020nwb} at
$\kappa=0.0749$ for ${\hat p}^2<0.1$. The lattice mass used here for
the rescaling was fixed at the value
$m_h=(2\kappa\chi)^{-1/2}=0.0769(8)$.} \label{susce}
\end{figure}

By computing the Fourier transformed connected propagator as
function of the lattice momentum $\hat{p}_\mu=2 \sin p_\mu/2$, the
extraction of $m^2_h$ is straightforward because its inverse is just
the zero-momentum propagator, or susceptibility $\chi$ \BE
2\kappa\chi\equiv 2\kappa G(p=0) = \frac{1}{m^2_h} \EE Instead, to
extract $M_h$  the propagator data were first fitted to the
2-parameter form \BE \label{twoparameter}
G_{\rm{fit}}(p)=\frac{Z_{\rm{prop}} }{{\hat p}^2 +m^2_{\rm{latt}}
}\EE in terms of the squared lattice momentum ${\hat p}^2$. The data
were then re-scaled by $({\hat p}^2+m^2_{\rm{latt}})$ so that
deviations from a flat plateau are immediately visible. This defines
the standard single-particle mass  and the difference from unity of
the height of the plateau, which is $Z_{\rm{prop}}$, measures the
residual coupling to multi-particle states.

The results in Fig.\ref{0.074} for $\kappa=0.074$, i.e. close to the
critical point but still in the symmetric phase, show that, there, a
single lattice mass works remarkably well in the whole range of
momentum down to $p=0$. Thus single-particle states and
multi-particles states are very weakly coupled in the whole momentum
range. At least with a lattice mass of about 0.2, one is then close
enough to the ``trivial'' continuum limit of $\Phi^4$ in 4D.

According to the standard point of view, in the broken phase there
should be no particular difference. However, it was already observed
in refs.\cite{Cea:1999kn} and \cite{Stevenson:2005yn} that, in this
case, the mass from the higher-momentum fit cannot describe the data
in the $p\to 0$ limit where the deviations from constancy become
highly significant. As a further check, one more simulation was
performed in ref. \cite{Consoli:2020nwb} on a large $76^4$ lattice
for $\kappa=0.0749$, which is even closer to the critical point. As
one can see from Fig.\ref{0933}, the mass parameter $m_{\rm{latt}}
\equiv M_h$, obtained from the fit to the propagator for ${\hat
p}^2>0.1$ cannot reproduce the data for ${\hat p}^2<0.1$. In this
low-momentum range, in fact, the data select a smaller mass, which
is very close to the inverse susceptibility
$m_h=(2\kappa\chi)^{-1/2}=0.0769(8)$, see Fig.\ref{susce}.

The difference between $m_h$ and $M_h$ determines zero-momentum
peaks, see Fig.\ref{0933}, which increase for $\kappa \to \kappa_c$.
The observed values, $M^2_h/m^2_h=$1.24(5), 1.31(5), 1.47(9),
respectively for $\kappa=$ 0.0751, 0.07504, 0.0749
\cite{Consoli:2020nwb}, are consistent with the expected logarithmic
trend $M^2_h= L m^2_h(c_2)^{-1}$ in Eq.(\ref{c2}), so that, as
anticipated in the Introduction, one can fit these data and obtain
$(c_2)^{-1/2} = 0.67 \pm 0.01 ~({\rm stat}) \pm 0.02 ~({\rm sys})$.
With this value, and the leading-order estimate $\lambda=
16\pi^2/(3L)$ in the relation $\lambda= 3 m^2_h/\langle \Phi
\rangle^2$, one finds $M_h = 690 \pm 10 ~({\rm stat}) \pm 20 ~({\rm
sys})~ {\rm GeV}$. Instead, with the next-to-leading $m_h-\langle
\Phi \rangle$ relation and the same $c_2$, we obtained $M_h = 750\pm
10 ~({\rm stat}) \pm 20 ~({\rm sys})$ GeV \cite{Consoli:2020nwb}.
The two values were then summarized into a final estimate $M_h \sim
720 \pm 30 $ GeV which accounts for the theoretical uncertainty and
updates the previous work of refs.\cite{Cea:2003gp, Cea:2009zc}.

\section{The role of $m_h$ and $M_h$ in the observable interactions}

The lattice simulations in the previous section, supporting the idea
of a scalar propagator which smoothly interpolates between a mass
scale $m_h$ and much larger $M_h$ with $M^2_h\sim L m^2_h$, are
consistent with the physical picture deduced from the effective
potential. To understand the interplay of the two masses and their
role in the observable processes, it is convenient to first follow
ref.\cite{Castorina:2007ng} where the phenomenology of a heavy but
weakly interacting Higgs resonance was first considered.

Differently from here, where $m_h$ and $M_h$ are assumed to coexist,
in ref.\cite{Castorina:2007ng} one was adopting the ideal
single-mass limit b) of the Introduction which (with the exception
of the Lorentz-invariant, zero-measure set $p_\mu=0$) effectively
reproduces a standard propagator with mass $M_h$. However, the
problem was the same considered here: a $\Lambda_s-$independent
scaling law $M_h=K \langle \Phi \rangle$. This opens a corner of
parameter space, namely large $K$ but $M_h \ll \Lambda_s$, that does
not exist in the conventional view. But, for this reason, the
constant $3K^2$ is now basically different from the coupling
$\lambda$ defined through the $\beta-$function \BE \label{beta}
\ln\frac{\mu}{\Lambda_s}=
\int^\lambda_{\lambda_0}~\frac{dx}{\beta(x)} \EE For  $\beta(x) =
3x^2/(16\pi^2) + O(x^3)$, whatever the contact coupling $\lambda_0$
at the asymptotically large $\Lambda_s$, at finite scales $\mu\sim
M_h$ this gives $\lambda\sim 16\pi^2/(3L)$ with $L= \ln
(\Lambda_s/M_h)$. As anticipated, this $\lambda$ (and not the
$\Lambda_s-$ independent $3K^2$) is the appropriate coupling to
describe the infinitesimal interactions of the fluctuations of the
broken phase.

Defining $M_w=g_{\rm gauge}\langle \Phi \rangle/2 $ and with the
notations of \cite{yndurain}, a convenient parametrization
\cite{Castorina:2007ng} of these residual interactions in the scalar
potential is ($r=M^2_h/4M^2_w = K^2/g^2_{\rm gauge}$) \BE
\label{potential} U_{\rm scalar}=\frac{1}{2}M^2_h h^2 + \epsilon_1 r
g_{\rm gauge} M_w h(\chi^a\chi^a + h^2) +\frac{1}{8} \epsilon_2 r
g^2_{\rm gauge} (\chi^a\chi^a + h^2)^2 \EE The two parameters
$\epsilon_1$ and $\epsilon_2$, which are usually set to unity,
account for $\lambda \neq 3K^2$, i.e. \BE \epsilon^2_1=\epsilon_2=
\frac{\lambda}{3K^2}\sim 1/L \EE But what about the full gauge
theory? At first sight, the original calculation \cite{lee} in the
unitary gauge could give the impression that $W_LW_L$ scattering is
indeed governed by the large coupling $ 3K^2 =3 M^2_h/\langle \Phi
\rangle^2$. To find the answer, let us recall the basics of that
calculation. One starts from a tree-level amplitude $A_0$ which is
formally $O(g^2_{\rm gauge})$ but ends up with \BE A_0(W_LW_L \to
W_LW_L)\sim \frac{3 M^2_h g^2_{\rm gauge} }{4 M^2_w} = 3K^2 \EE Here
the factor $g^2_{\rm gauge}$ comes from the vertices. The $1/M^2_w$
derives from the external longitudinal polarizations
$\epsilon^{(L)}_\mu \sim (k_\mu/M_w)$ and $M^2_h$ emerges after
expanding the propagator \BE \frac{1 }{s- M^2_h}\to \frac{1 }{s}(1 +
\frac{M^2_h }{s}+...) \EE While the leading $1/s$ contribution
cancels against a similar term from the other diagrams (which
otherwise would give an amplitude growing with $s$), the $M^2_h$
from the 2nd-order term is effectively ``promoted'' to coupling
constant reproducing the same result of a pure $\Phi^4$ with contact
coupling $\lambda_0=3K^2$ at some large scale $\Lambda_s$.

However, in $\Phi^4$, this is just the tree approximation with the
same coupling at all momentum scale. To find the $\chi\chi \to \chi
\chi$ amplitude at some scale $\mu$, we should instead use
Eq.(\ref{beta}) and let the coupling evolve, from $\lambda_0$ to
$\lambda$, as previously done in Eq.(\ref{potential}), i.e. \BE
A(\chi\chi \to \chi \chi)\Big|_{g_{\rm gauge}=0}\sim \lambda \sim
\frac{1 }{\ln (\Lambda_s/\mu)}\EE Thus, recalling that the
Equivalence Theorem is valid to all orders in the scalar
self-interactions but to lowest order in $g_{\rm gauge}$
\cite{Cornwall:1974km,Chanowitz:1985hj,Bagger:1989fc}, we obtain the
result anticipated in the Introduction \BE A(W_LW_L \to W_LW_L)= [1
+O(g^2_{\rm gauge})]~A(\chi\chi \to \chi \chi)\Big|_{g_{\rm
gauge}=0}=O(\lambda)\EE This analysis, from our present perspective
where $m_h$ and $M_h$ coexist and could be experimentally
determined, shows that at $\mu \sim M_h$ the supposed strong
interactions proportional to $\lambda_0= 3K^2$ are actually
controlled by the much smaller coupling \BE \lambda= \frac{3m^2_h
}{\langle \Phi \rangle^2 }= 3K^2 ~\frac{m^2_h }{M^2_h}\sim 1/L \EE
Analogously, the conventional very large width into longitudinal
vector bosons computed with $\lambda_0= 3K^2$, say $\Gamma^{\rm
conv}(M_h \to W_LW_L) \sim M^3_h/\langle \Phi \rangle^2$, should
instead be rescaled by $\epsilon_1^2= (\lambda/3K^2)=m^2_h/M^2_h$.
This gives \BE \label{aux} \Gamma(M_h\to W_LW_L) \sim \frac{m^2_h
}{M^2_h} ~\Gamma^{\rm conv}(M_h \to W_LW_L) \sim M_h ~\frac{m^2_h
}{\langle \Phi \rangle^2} \EE where $M_h$ indicates the available
phase space in the decay and ${m^2_h }/{\langle \Phi \rangle^2} $
the interaction strength. Therefore, it is through the decays of the
heavier state that the coupling $\lambda= 3m^2_h/\langle \Phi
\rangle^2$ could become visible, thus confirming that $m_h$ and
$M_h$ represent excitations of the same field.

\section{Comparison with 4-lepton data}

Suppose to take seriously the idea of a second heavier excitation of
the Higgs field with mass $M_h\sim$ 700 GeV. Are there experimental
indications for such resonance? Furthermore, what kind of
phenomenology should we expect? Finally, what about the
identification $m_h\sim$ 125 GeV, implicitly assumed for our lower
mass scale? In the following, I will summarize the results of
ref.\cite{CC2020} where these questions were addressed in connection
with a certain excess of 4-lepton events observed by ATLAS
\cite{Aaboud:2017rel,ATLAS2} for invariant mass $\mu_{4l}\sim $ 700
GeV ($l=e,\mu$).

Of course, the 4-lepton channel is just one possible decay channel
and, for a comprehensive analysis, one should also look at the other
final states. For instance, at the 2-photon channel that, in the
past, has been showing some intriguing signal for the close energy
of 750 GeV. However, the 4-lepton channel is experimentally clean
and, for this reason, is considered the ``golden'' channel for a
heavy Higgs resonance. Moreover, the bulk of the effect can be
analyzed at an elementary level. Thus it makes sense to start from
here.

The main new aspect is the strong reduction of the conventional
width in Eq.(\ref{aux}). For $M_h=$ 700 GeV, where $ \Gamma^{\rm
conv}( M_h \to ZZ)\sim 56.7~{\rm GeV}$
\cite{Djouadi:2005gi,handbook}, fixing $m_h=125$ GeV gives \BE
\Gamma( M_h \to ZZ)\sim \frac{m^2_h}{(700~{\rm GeV})^2}~56.7~{\rm
GeV}\sim 1.8~{\rm GeV} \EE Afterward, by maintaining the other
contributions for $M_h=$ 700 GeV \cite{Djouadi:2005gi,handbook} \BE
\Gamma(M_h\to {\rm fermions+ gluons+ photons...})\sim 28~{\rm
GeV}\EE and with the same ratio $ \Gamma( M_h \to WW)/\Gamma( M_h
\to ZZ)\sim$ 2.03, we find a total width \BE \label{th5}\Gamma( M_h
\to all) \sim 28~{\rm GeV} + 3.03~\Gamma (M_h \to ZZ) \sim 33.5~{\rm
GeV} \EE and a fraction $ B( M_h \to ZZ)\sim (1.8~/~33.5) \sim
0.054$.

Now, the production cross section $\sigma (pp\to M_h)$. Here the
main contributions are the basic Gluon-Gluon Fusion (GGF) and
Vector-Boson Fusion (VBF) processes, where two gluons or two vector
bosons $VV$ ($WW$ or $ZZ$) fuse to produce the heavy state $M_h$,
i.e. \BE \sigma (pp\to M_h)\sim \sigma (pp\to M_h)_{\rm GGF}+ \sigma
(pp\to M_h)_{\rm VBF} \EE For the GGF term I will consider two
estimates: $\sigma (pp\to M_h)_{\rm GGF}=800(80)$ fb from
ref.\cite{Djouadi:2005gi} and $\sigma (pp\to M_h)_{\rm
GGF}=1078(150)$ fb from ref.\cite{handbook}. These refer to
$\sqrt{s}=$ 14 TeV and should be rescaled by about $-12\%$ for
$\sqrt{s}=$ 13 TeV.

\begin{table*}
\caption{\it For $M_h=$ 700 GeV and $m_h=$ 125 GeV, we report our
predictions for the peak cross section $\sigma_R(pp\to 4l)$ and the
number of events at two values of the luminosity. The two total
cross sections are our extrapolation to $\sqrt{s}=$ 13 TeV of the
values in \cite{Djouadi:2005gi} and \cite{handbook} for $\sqrt{s}=$
14 TeV. As explained in the text, only the GGF mechanism is relevant
in our model. }
\begin{center}
\begin{tabular}{cccc}
~~~~~~~~$\sigma(pp\to M_h)$ ~~~~~~~~~~~~&  $\sigma_R(pp\to 4l)$   ~~~~~~~& $n_{\rm 4l}$ (36.1$fb^{-1})$ ~~~~&$n_{\rm 4l}$(139$fb^{-1})$ \\
\hline
700(70)~fb   &  0.17(2)~fb  &$6.0\pm 0.6$  &  $23.2\pm 2.3$       \\
\hline
950(150)~fb   & 0.23(4)~fb  & $8.1\pm 1.3$ & $31.5 \pm 5.0$             \\
\hline
\end{tabular}
\end{center}
\end{table*}

About VBF, I observe that the $VV\to M_h$ process is the inverse of
the $M_h\to VV$ decay so that $\sigma(pp\to M_h)_{\rm VBF}$ can be
expressed \cite{kane} as a convolution with the parton densities of
the same Higgs resonance decay width. Thus with a large $3K^2$
coupling to longitudinal $W$'s and $Z$'s and conventional width
$\Gamma^{\rm conv}(M_h \to WW+ZZ) \sim$ 172 GeV, the VBF mechanism
would become sizeable. But this coupling is not present in our
picture, where instead $\Gamma(M_h \to WW+ZZ) \sim$ 5.5~{\rm GeV}.
Therefore, the VBF will correspondingly be reduced from its
conventional estimate $\sigma^{\rm conv}(pp\to M_h)_{\rm
VBF}=250\div300$ fb by the small ratio (5.5 / 172)$\sim$ 0.032. This
gives $\sigma(pp\to M_h)_{\rm VBF}~ \lesssim 10~ $ fb and can be
neglected.

In the end, for a relatively narrow resonance the effects of its
virtuality should be small. Thus one can approximate the resonance
cross section by on-shell branching ratios as \BE\label{exp3}
\sigma_R (pp\to 4l)\sim \sigma (pp\to M_h)\cdot B( M_h \to ZZ) \cdot
4 B^2( Z \to l^+l^-) \EE Altogether, for $B( M_h \to ZZ)\sim $ 0.054
and $4B^2( Z \to l^+l^-)\sim 0.0045$, the expected peak cross
section and numbers of events (for efficiency $\sim$ 0.98) are
reported in Table 1. \vskip 17pt

\begin{figure}[ht]
\begin{center}
\psfig{figure=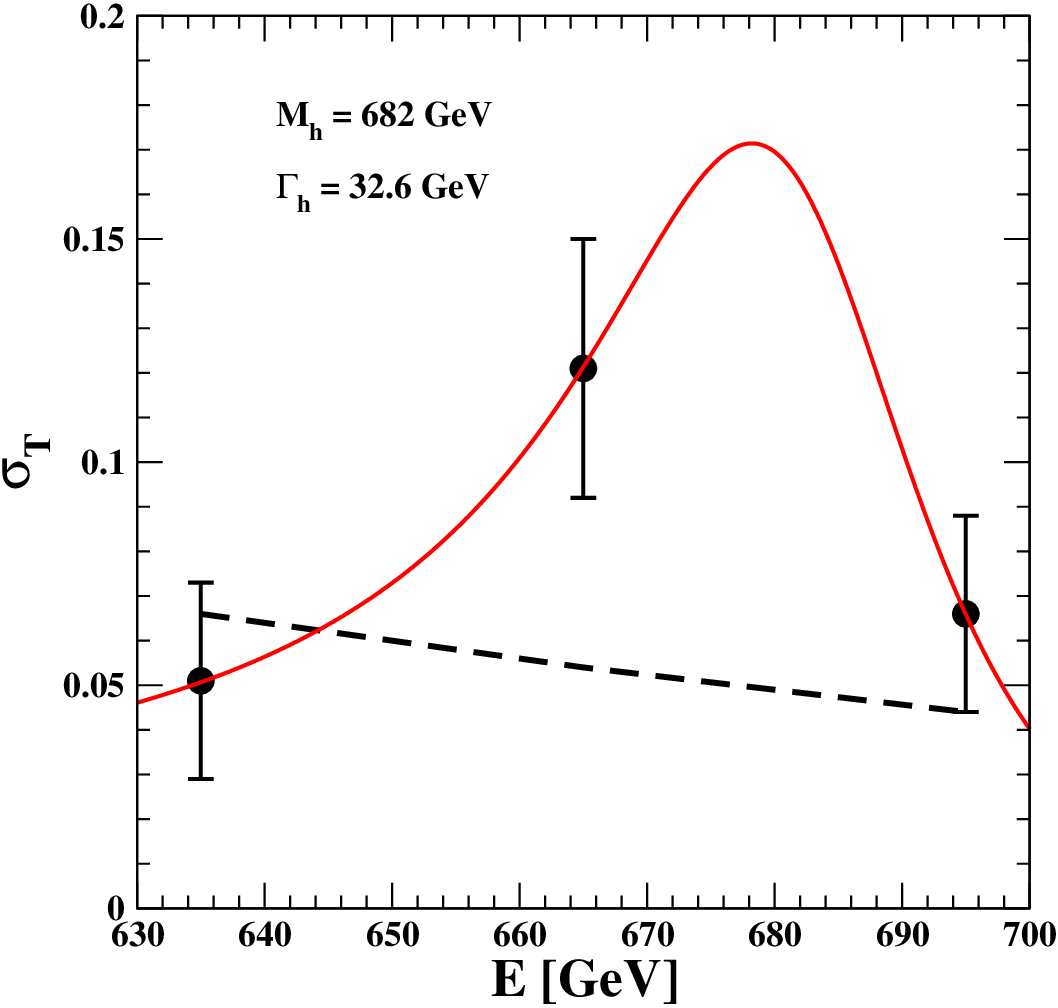,width=6.5truecm,angle=0}
\end{center}
\caption{ {\it For $M_h=$ 682 GeV and $\Gamma_h=$ 32.6 GeV, a fit
(in red) with Eq.(\ref{sigmat}) to the ATLAS data converted into
cross sections in fb (the black dots). The fitted parameters are
$\sigma_R = 0.154\pm 0.054$ fb and
$\sigma_B=0.009^{+0.017}_{-0.008}$ fb, or total fitted background
$N^{\rm fit}_B= 4^{+7}_{-4}$. Errors from fit give $ \chi^2=$ 2.41
with $\chi^2_{\rm min}=0$. The dashed line is ATLAS background
estimate.}} \label{fit}
\end{figure}

\vskip 2pt
\begin{figure}[ht]
\begin{center}
\psfig{figure=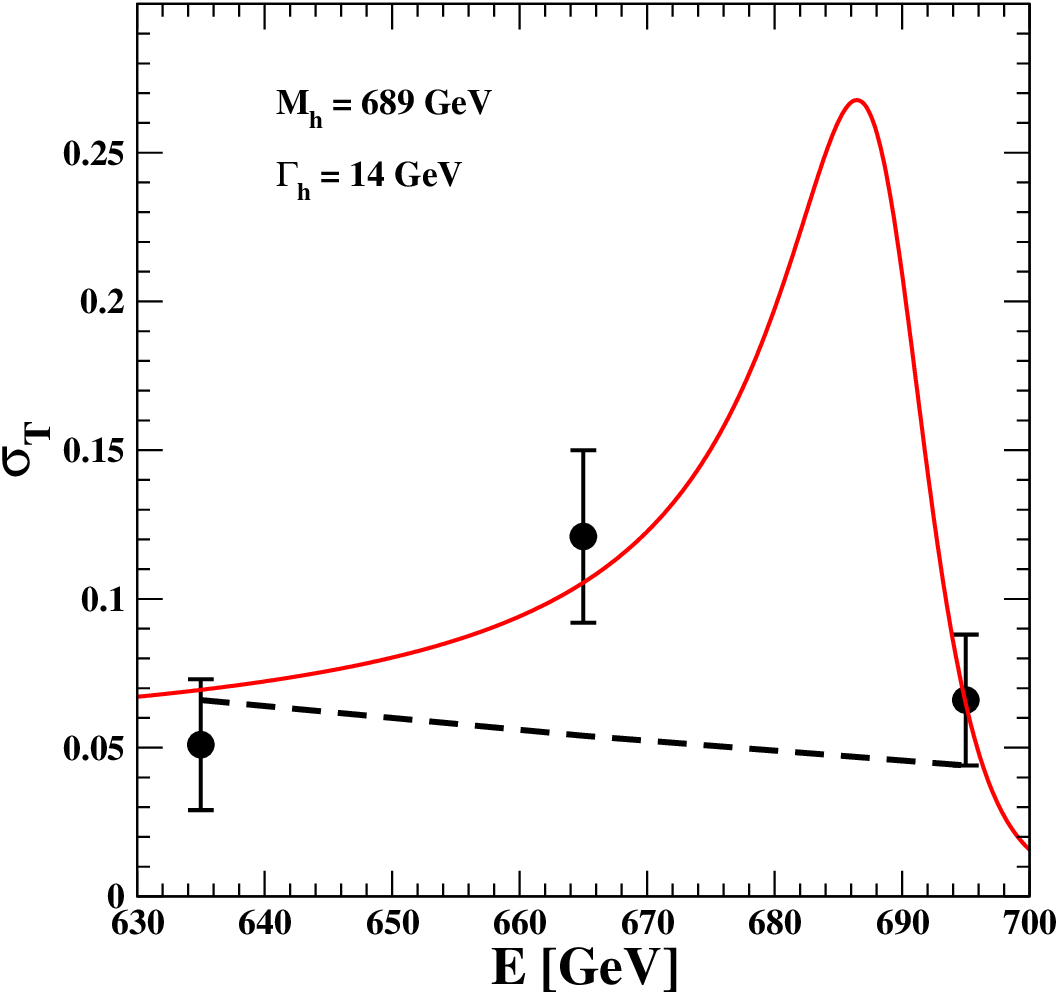,width=6.5truecm,angle=0}
\end{center}
\caption{ {\it As in Fig.\ref{fit} for $M_h=$ 689 GeV and
$\Gamma_h=$ 14 GeV. Fitted parameters are $\sigma_R =
0.19^{+0.16}_{-0.13}$ fb and $\sigma_B=0.04^{+0.01}_{-0.03}$ fb, or
total fitted background $ N^{\rm fit}_B=16^{+4}_{-12}$. Errors from
fit give $\chi^2=$ 2.41 with $\chi^2_{\rm min}=1$. The dashed line
is ATLAS background estimate.}} \label{fit2}
\end{figure}

For the smaller statistics of 36.1 fb$^{-1}$, see Fig.4a of
\cite{Aaboud:2017rel}, these predictions can be directly compared,
and are well consistent, with the measured value $ n_{\rm 4l}\sim
(6\pm 3)$ for $\mu_{\rm 4l}=$ 700 GeV. Instead, to compare with the
larger ATLAS sample \cite{ATLAS2} of 139 fb$^{-1}$ a different
treatment is needed. In fact, in their Fig.2d there are now {\it
three} points in the relevant energy region: at $\mu_{\rm 4l}\sim
635(15)$ GeV, where $ n_{\rm 4l}\sim 7.0\pm 3.0 $, at $\mu_{\rm
4l}\sim 665(15)$ GeV, where $ n_{\rm 4l}\sim 16.5\pm 4.0 $, and at
$\mu_{\rm 4l}\sim 695(15)$ GeV, where $ n_{\rm 4l}\sim
9.0^{+2.2}_{-3.0} $. Therefore, by defining $\mu_{4l}= E$ and
$s=E^2$, we have assumed that these 4-lepton events derive from the
interference of a resonating amplitude $A^{R}(s)\sim 1/(s - M^2_R)$
and a slowly varying background $A^{B}(s)$. For positive
interference below peak, setting $M^2_R=M^2_h -i M_h \Gamma_h$, this
gives a total cross section \BE \label{sigmat} \sigma_T=\sigma_B
-\frac{2(s-M^2_h)~\Gamma_h M_h}{(s-M^2_h)^2+ (\Gamma_h
M_h)^2}~\sqrt{\sigma_B\sigma_R} +\frac{(\Gamma_h M_h)^2
}{(s-M^2_h)^2+ (\Gamma_h M_h)^2}~ \sigma_R\EE  where, in principle,
both the average background $\sigma_B$ and the resonating $\sigma_R$
could be treated as free parameters. By converting event numbers
into cross sections, the best fit is at $M_h=$ 682 GeV, see
Fig.\ref{fit}. Since in our model, for small changes of the mass,
$\Gamma_h$ varies linearly with $M_h$, the width was constrained by
the relation $(\Gamma_h/M_h)=(33.5/700)$. ATLAS estimate of the
background is also shown as a dashed line.

\begin{figure}[ht]
\begin{center}
\psfig{figure=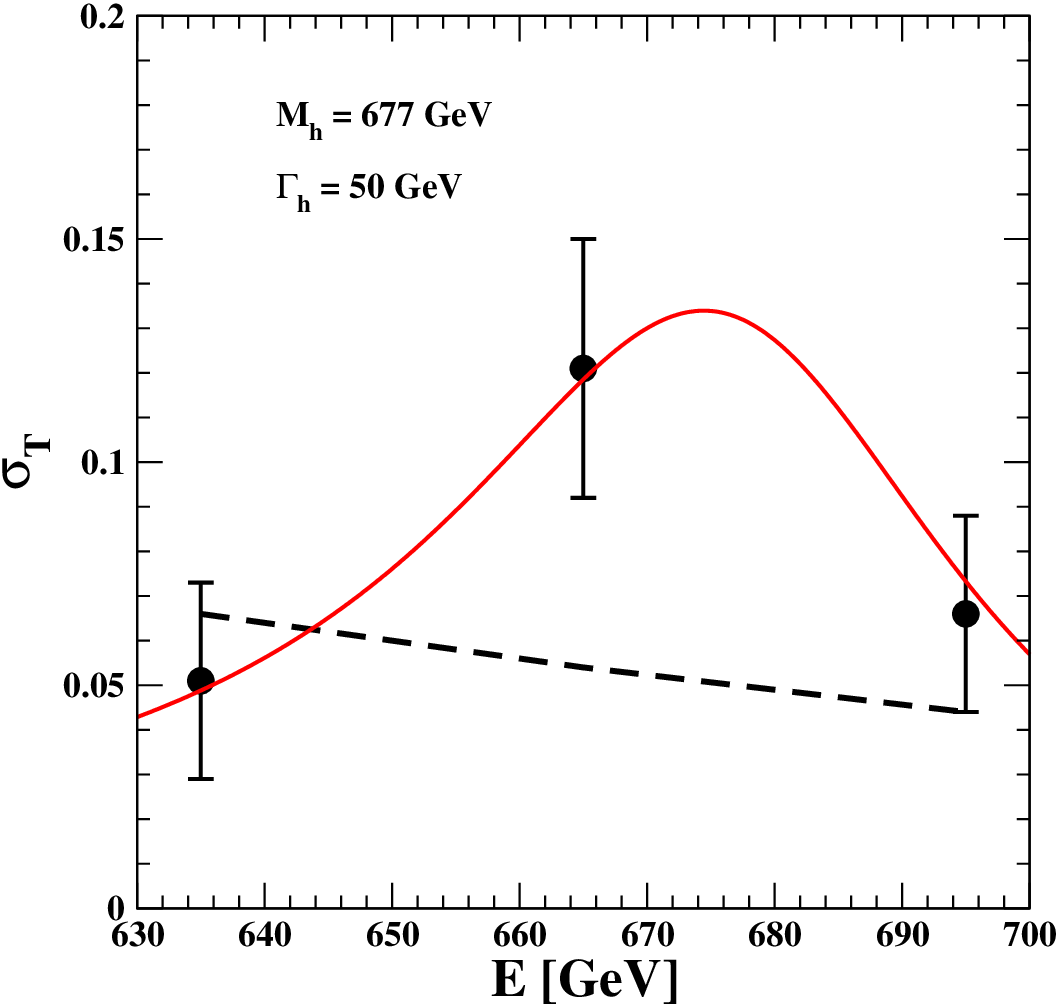,width=6.5truecm,angle=0}
\end{center}
\caption{ {\it As in Fig.\ref{fit} for $M_h=$ 677 GeV and
$\Gamma_h=$ 50 GeV. Fitted parameters are $\sigma_R = 0.131\pm
0.036$ fb and $\sigma_B=0.002^{+0.012}_{-0.002}$ fb, or total fitted
background $ N^{\rm fit}_B=1^{+5}_{-1}$. Errors from fit give $
\chi^2=$ 2.41 with $\chi^2_{\rm min}=0.1$. The dashed line is ATLAS
background estimate.}} \label{fit3}
\end{figure}

With this theoretical input the fit yields a small average
background $\sigma_B=0.009^{+0.017}_{-0.008}$ fb, equivalent to
total background events $N^{\rm fit}_B= 4^{+7}_{-4}$, and a
$\sigma_R$ consistent with Table 1. Therefore, if the observed
number of events $N_{\rm obs}\sim 32 \pm 7$ is expressed as $N_{\rm
obs}= \Delta N + N_B$ the fit would give a minimum number of extra
non-background events $(\Delta N)^{\rm fit} \sim 21 \pm 7$, i.e. a
genuine non-zero signal at the 3-sigma level. To better understand
the relation between strength of the signal and resonance width, I
have performed other fits by first decreasing the width. This can
give a background closer to ATLAS estimate $N_B({\rm ATLAS})\sim$
22. In fact, an acceptable fit is still obtained with $\Gamma_h=$ 14
GeV, $M_h=$ 689 GeV and a larger fitted background $ N^{\rm
fit}_B=16^{+4}_{-12} $, see Fig.\ref{fit2}. At the opposite side a
good fit can also be obtained for $\Gamma_h=$ 50 GeV and $M_h=$ 677
GeV, see Fig.\ref{fit3}, but, in this case, the fitted number of
background events $N^{\rm fit}_B= 1^{+5}_{-1}$ is much, much lower
than the ATLAS estimate.

\vskip 15pt

Conclusion: with Eq.(\ref{sigmat}) and the average background as
free parameter, one finds an excellent fit of the ATLAS data, see
Fig.\ref{fit}, with our reference $\Gamma_h/M_h$ ratio, a $\sigma_R$
as in Table 1 and a small background. Acceptable fits can however be
obtained with smaller widths and a background closer to ATLAS value,
see Fig.\ref{fit2}. Altogether, the observed range of the various
parameters can be approximated as: $M_h \sim (680\pm 10)$ GeV,
$\Gamma_h \sim (32\pm 18)$ GeV, $\sigma_R \sim (0.15\pm 0.05)$ fb,
$\sigma_B \sim 0.01^{+0.03}_{-0.01}$ fb. In particular, the fitted
$M_h$ stays in the theoretical band $M_h=690\pm 10 ~({\rm stat}) \pm
20 ~({\rm sys})$ GeV obtained when our lattice data
\cite{Consoli:2020nwb} are combined with the leading-order
$m_h-\langle\Phi\rangle$ relation.

Therefore, for the special role of the 4-lepton channel, further
checks of the background and further statistical tests are needed.
For instance, with the 36.1 fb$^{-1}$ luminosity, the deep diving of
the local $p_0$ at the 3.6 sigma level \cite{denys} was already
suggesting a new narrow resonance at 700 GeV. It remains to be seen
if an unambiguous answer could still be obtained with the present
LHC configuration \footnote{To this end, 4-lepton data from CMS
would be crucial. At present there are no data with the full
statistics of 139 fb$^{-1}$ but partial results are given in
previous reports. For instance, for 12.9 fb$^{-1}$, in Fig.3 (right)
of  CMS PAS HIG-16-033 of 2016/08/04, at $\mu_{\rm 4l}=660(10)$ GeV,
the event number $n_{\rm 4l} = 4^{+3}_{-2}$ is much larger than the
estimated background of 0.4$\div$0.5 events. This would give
$\sigma_R(pp\to 4l)\sim (0.3 \pm 0.2)$ fb, consistently with our
Table 1. More extensive data were reported in CMS PAS HIG-18-001 of
2018/06/03 but the very compressed scale, see Figs.2 (left) and 9,
prevents to extract the numerical values in the relevant region
around 680 GeV. Finally, in Fig.6 of CMS PAS HIG-19-001 of
2019/03/22, the data stop at $\mu_{\rm 4l}=500$ GeV.} or must be
postponed to the high-luminosity phase .

\section{Summary and outlook}

After the observation of the narrow scalar resonance at 125 GeV,
Veltman's original idea of a naturally large Higgs particle mass
seems to be ruled out. Yet, perhaps, the last word has not been
said. If SSB in $\Phi^4$ theories is really a weak first-order phase
transition, as indicated by most recent lattice simulations, one
should consider approximations to the effective potential which are
consistent with this scenario. Then, by combining analytic
calculations and lattice simulations, it becomes conceivable that,
besides the known resonance with mass $m_h\sim$ 125 GeV, a new
excitation with mass $M_h\sim 700$ GeV might show up. The
peculiarity, though, is that the heavier state should couple to
longitudinal vector bosons with the same typical strength of the
low-mass state and would thus represent a relatively narrow
resonance. In this case, such hypothetical new resonance would
naturally fit with some excess of 4-lepton events observed by ATLAS
around 680 GeV. Analogous data from CMS are needed to definitely
confirm or disprove this interpretation.

But, before concluding, I will discuss the possible implications
that a two-mass structure of the Higgs field may have for radiative
corrections. Our lattice simulations in Sect.4 indicate two regimes
of the inverse propagator $G^{-1}(p)$. This behaves as $(p^2 +
m^2_h)$ in the low-$p^2$ limit, see Fig.\ref{susce} for $p^2<$ 0.1,
and as $(p^2 + M^2_h)$ at larger $p^2$, see Fig.\ref{0933} for $p^2
>$ 0.1. Extrapolating
the observed scaling $M^2_h\sim m^2_h\ln (\Lambda_s/M_h)$ to very
large values of $\Lambda_s$ gives the idea, in Minkowski space, of
two vastly different mass-shell regions, as when the spectral
density has not the standard single-peak structure.

By expressing $G^{-1}(p)= p^2 - \Pi(p)$, the observed $m_h-M_h$
difference confirms the point of view of the Introduction that, in
the broken-symmetry phase, the self-energy function $\Pi(p)$
exhibits a non-trivial momentum dependence. This can be interpreted
as the coexistence, in the cutoff theory, of two kinds of
quasi-particles associated respectively with two distinct aspects of
the effective potential: its quadratic shape and the zero-point
energy which determines its depth. The analogy with superfluid He-4,
where the observed energy spectrum arises by combining the two
quasi-particle spectra of phonons and rotons, would then suggest a
model propagator \BE \label{interpol} G(p)= \frac{1 -
I(p)}{2}\frac{Z_1}{p^2 + m^2_h }+\frac{1 + I(p)}{2}\frac{Z_2}{p^2 +
M^2_h } \EE where the interpolating function $I(p)$ depends on an
intermediate momentum scale $p_0$ and tends to $+1$ for large
$p^2\gg p^2_0$ and to $-1$ when $p^2 \to 0$. For instance, by fixing
sharply the central values of Sect.4, $Z_1=$ 0.945, $m_h=$ 0.0769,
$Z_2=$ 0.966, $M_h=$ 0.0933, the form $I(p)= \tanh
(p^2-p^2_0)/\gamma$ gives a good fit to the lattice data, $\chi^2=
29/ (40 -6)$, for $p^2_0=0.087$ and $\gamma=0.0043$. But small
changes of the $Z$'s and of the masses can induce sizeable changes
of $p^2_0$ and $\gamma$ indicating that the crossover region may be
wider than with a simple step function. Moreover, any trial form for
$I(p)$ in Eq.(\ref{interpol}) introduces a model dependence that
could obscure the significance of the results. Thus, while the
lattice data indicate that, in Minkowski space, the spectral density
should exhibit a two-peak structure, in practice, performing the
analytic continuation in a non-perturbative regime, and in a
numerical simulation where only a discrete set of data is available,
is a difficult task \cite{spectral}.

To get some intuitive insight on the interplay of the two masses in
radiative corrections, I will thus refer to a work of van der Bij
\cite{jochum2018} where a propagator which resembles
Eq.(\ref{interpol}) was also considered. To this end, he starts from
two observations. First, renormalizability, by itself, does not mean
a single-particle peak but only a spectral density which falls off
sufficiently fast at infinity. Second, the Higgs field fixes the
vacuum state of the theory which determines the masses of all other
particles. Therefore, the Higgs field itself remains different and
it is not unreasonable to expect a spectral density which is not a
single $\delta-$function. Here, he does not mention the two-branch
spectrum of superfluid He-4 but the idea of the SSB vacuum as some
kind of medium seems implicit in this remark. With a definite
example \cite{hill}, he then considers the possibility that the
physical Higgs boson is actually a mixture of two states with a
spectral density approximated by two $\delta-$function peaks. The
resulting propagator structure can be written as ($-1\leq\eta\leq
1$) \BE \label{mixing} G(p) \sim \frac{1 - \eta}{2} \frac{1}{p^2 +
m^2_h } + \frac{1 + \eta}{2} \frac{1}{p^2 + M^2_h } \EE and could be
used in the analysis of the $\rho-$parameter \cite{rho1,rho2}. Since
the two-loop correction \cite{jochumtini} is completely negligible
for masses below 1 TeV, one can restrict to the one-loop level,
where the two branches Eq.(\ref{mixing}) do not mix, as when
replacing in the main logarithmic term an effective mass $m_{\rm
eff} \sim \sqrt{m_h M_h}~(M_h/m_h)^{\eta/2}$. In our case, this
would be between $m_h=$ 125 GeV and $M_h\sim $ 700 GeV so that it
becomes important to understand how well the mass parameter obtained
indirectly from radiative corrections agrees with the $m_h=$ 125
GeV, measured directly at LHC.

Here, after the (partial) reassessment of the NuTeV anomaly
\cite{nutev,partial}, only two measurements give sharp indications.
Namely, $A_{\rm FB}(b)$ which favors a large effective mass and
$A_{\rm LR}$ from SLD which goes in the opposite direction. This is
well illustrated in the PDG review \cite{Tanabashi:2018oca}. In
fact, from the experimental set ($A_{\rm LR}$, $M_Z$, $\Gamma_Z$,
$m_t$), one would predict the pair $[m_{\rm eff}=38^{+30}_{-21} $
GeV, $\alpha_s(M_z)=0.1182(47)$]. While, from the set ($A_{\rm
FB}(b,c)$, $M_Z$, $\Gamma_Z$, $m_t$), the other pair [$m_{\rm
eff}=348^{+187}_{-124}$ GeV, $\alpha_s(M_z)=0.1278(50)$].

These two extreme cases show that, at this level of precision, we
should try to evaluate the uncertainty induced by strong
interactions. This enters indirectly, for instance in the $M_w-M_z$
interdependence, through the contribution of the hadronic vacuum
polarization to $\Delta \alpha(M_z)$, but also directly through the
value of $\alpha_s(M_z)$. More precisely, in the two examples
considered above, this uncertainty enters through $r(M_z)$, the
strong-interaction correction to the quark-parton model in
$\sigma(e+e^- \to hadrons)$ at center of mass energy $Q=M_z$.
Through the total W and Z widths, and the LEP1 peak cross sections,
this affects all quantities, even the pure leptonic widths and
asymmetries.

Since a small coupling does not guarantee, by itself, a good
convergence of the perturbative expansion, one should seriously
consider that, even at large center of mass energies, the
experimental quantity $ r^{\rm EXP}(Q)$ obtained from the data can
sizeably differ from its theoretical prediction $ r^{\rm TH}(Q)= 1 +
\alpha_s(Q)/\pi + ...$ computed from the first few terms. In this
case, for a real precision test, instead of treating $\alpha_s(M_z)$
as a free parameter, one could extrapolate $ r^{\rm EXP}(Q)$ toward
$Q=M_z$ and use this value to extract the EW corrections from
experiments.

As pointed out in ref.\cite{fiore1992}, in fact, there is some
excess in the data so that, to extrapolate correctly from PETRA, PEP
and TRISTAN toward the $Z$ peak, one should replace in $ r^{\rm TH}$
(34 GeV) a considerably larger $\alpha_s$(34 GeV)$\sim $ 0.17
instead of the canonical 0.14 predicted from deep inelastic
scattering. This $\Delta \alpha_s=+0.03$ is a small $+1\%$ effect in
the QCD correction but is visible in the slope of the $\gamma-Z$
interference. On the $Z$ peak, the effect is smaller because we are
now speaking of a shift from $\alpha_s(M_z) = 0.118$ to
$\alpha_s(M_z) = 0.128$ which is just a $+0.3\%$ effect in $r^{\rm
TH}(M_z)$. Nevertheless, the Higgs mass parameter extracted from the
LEP1 data would be considerably increased \cite{hioki1,hioki2}.

Later on, some excess in the total hadronic cross section had also
been observed at LEP2 \cite{erler,wilkinson,wynhoff} so that the
whole issue of $\sigma(e+e^- \to hadrons)$ was reconsidered by
Schmitt \cite{schmitt} in a thorough analysis of all data in the
range 20 GeV$\leq Q\leq$ 209 GeV. His conclusion was that,
individually, none of the measurements shows a significant
discrepancy. However, when taken together, there is an overall
excess at the 4-sigma level. If translated into the QCD correction,
this corresponds to replacing the higher range of values
$\alpha_s(M_z)\gtrsim$ 0.128 in $r^{\rm TH}(M_z)$ and, if used to
evaluate the EW corrections, would increase the value of $m_{\rm
eff}$ obtained from many experimental quantities. For instance, from
the set ($A_{\rm LR}$, $A_{\rm FB}(b,c)$, $M_Z$, $\Gamma_Z$, $m_t$).

In this sense, the present view, that the Higgs mass parameter
extracted indirectly from radiative corrections agrees perfectly
with the $m_h=$ 125 GeV measured directly at LHC, is not free of
ambiguities and one could in the end discover other motivations for
a new resonance, quite independently of the effective potential
and/or of lattice simulations of the propagator. This emphasizes
once more the importance of new, combined LHC measurements, starting
from the ``golden'' 4-lepton channel around 700 GeV.


\centerline{\bf DEDICATION} \vskip 5 pt

\noindent This paper is dedicated to the memory of Professor M. J.
G. Veltman. Perhaps now, that he is no longer with us, we can better
realize how much his moral legacy has transcended the purely
scientific.





\begin{thebibliography}{24}


\bibitem{veltman1977}
M. Veltman, Acta Phys. Polon. B\textbf{8} (1977) 475.

\bibitem{Englert:1964et}
F.~Englert, R.~Brout, Phys. Rev. Lett. \textbf{13} (1964) 321.

\bibitem{Higgs:1964ia}
P.W. Higgs, Phys. Lett. \textbf{12} (1964) 132.

\bibitem{glashow1961}
S. L. Glashow, Nucl. Phys. B\textbf{22} (1961) 579.


\bibitem{Aad:2012tfa}
ATLAS Collaboration (G.~Aad {\em et~al.}), Phys. Lett. B{\bf 716}
(2012) 1.


\bibitem{Chatrchyan:2012xdj}
CMS Collaboration (S.~Chatrchyan {\em et~al.}), Phys. Lett. B
\textbf{716} (2012) 30.

\bibitem{Coleman:1973jx}
S.R. Coleman, E.J. Weinberg, Phys. Rev. D\textbf{7} (1973) 1888.

\bibitem{Consoli:2020nwb}
M.~Consoli, L.~Cosmai, Int. J. Mod. Phys. A \textbf{35} (2020)
2050103, hep-ph/2006.15378.

\bibitem{symmetry}
M. Consoli, L. Cosmai, Symmetry {\bf 12}, 2020, 2037;
doi:10.3390/sym12122037.


\bibitem{lundow2009critical}
P.H. Lundow, K.~Markstr{\"o}m, Physical Review E \textbf{80} (2009)
031104.

\bibitem{Lundow:2010en}
P.H. Lundow, K.~Markstr{\"o}m, Nucl. Phys. B\textbf{845} (2011) 120.


\bibitem{akiyama2019phase}
S.~Akiyama, et al.,  Phys. Rev. D
  \textbf{100} (2019) 054510.


\bibitem{Cea:2003gp}
P.Cea, M.Consoli, L.Cosmai, Nucl.Phys.Proc.Suppl. \textbf{129}
(2004)780, hep-lat/0309050.

\bibitem{Cea:2009zc}
P.~Cea, L.~Cosmai, ISRN High Energy Phys. (2012) 637950,
hep-ph/0911.5220.

\bibitem{Consoli:1993jr}
M.~Consoli, P.M. Stevenson, Z. Phys. C \textbf{63} (1994) 427,
hep-ph/9310338.


\bibitem{Consoli:1997ra}
M.~Consoli, P.M. Stevenson, Phys. Lett. B \textbf{391} (1997) 144.

\bibitem{Consoli:1999ni}
M.~Consoli, P.M. Stevenson, Int. J. Mod. Phys. A \textbf{15}(2000)
133, hep-ph/9905427.


\bibitem{jochum2018}
J. J. van der Bij, Acta Phys. Polon. B\textbf{11} (2018) 397.

\bibitem{Tanabashi:2018oca}
M.~Tanabashi et~al. (Particle Data Group), Phys. Rev. D \textbf{98}
(2018) 030001.



\bibitem{shirkov}
D. Shirkov, Lectures at KEK, April 1991, KEK-91-13, 1992
(unpublished).

\bibitem{chetyrkin}
K. G. Chetyrkin, S. G. Gorishny, S. A. Larin, and F. V. Tkachov,
Phys. Lett. B  {\bf 132}(1983) 351.

\bibitem{kazakov}
D. I. Kazakov, Phys. Lett. B {\bf 133} (1983) 406.

\bibitem{stevenson1987}
P. M. Stevenson, Z. Phys. C{\bf 35} (1987) 467.

\bibitem{mpla1996}
M. Consoli and P. M. Stevenson, Mod. Phys. Lett. A {\bf 11} (1996)
2511.


\bibitem{Barnes:1978cd}
T.~Barnes and G.~I. Ghandour, Phys. Rev. D{\bf 22} (1980) 924.

\bibitem{Stevenson:1985zy}
P.~M. Stevenson, Phys. Rev. D{\bf 32} (1985) 1389.

\bibitem{Stancu:1989sk}
I.~Stancu and P.~M. Stevenson, Phys. Rev. D{\bf 42} (1990) 2710.


\bibitem{Cea:1996pe}
P.~Cea and L.~Tedesco, Phys. Rev. D{\bf 55} (1997) 4967.





\bibitem{Stevenson:2005yn}
P.~M. Stevenson, Nucl. Phys. B{\bf 729} (2005) 542.



\bibitem{Cea:1999kn}
P.~Cea, M.~Consoli, L.~Cosmai and P.~M. Stevenson, Mod. Phys. Lett.
A{\bf 14} (1999) 1673.




\bibitem{Castorina:2007ng}
P.~Castorina, M.~Consoli, D. Zappal\`a, J. Phys. G \textbf{35}
(2008) 075010, hep-ph/0710.0458.

\bibitem{yndurain}
M. J. G. Veltman and F. Yndurain, Nucl. Phys. B \textbf{325} (1989)
1.


\bibitem{lee}
B. W. Lee, C. Quigg, H. B. Tacker, Phys. Rev. D \textbf{16} (1977)
1519.



\bibitem{Cornwall:1974km}
J.M. Cornwall, D.N. Levin, G.~Tiktopoulos, Phys. Rev. D \textbf{10}
(1974) 1145.

\bibitem{Chanowitz:1985hj}
M.S. Chanowitz, M.K. Gaillard, Nucl. Phys. B \textbf{261} (1985)
379.

\bibitem{Bagger:1989fc}
J.~Bagger and C.~Schmidt, Phys. Rev. D \textbf{41} (1990) 264.

\bibitem{CC2020}
M. Consoli and L. Cosmai, arXiv:2007.10837v2 [hep-ph], 25/12/2020.

\bibitem{Aaboud:2017rel}
M.~Aaboud et~al. (ATLAS), Eur. Phys. J. C \textbf{78} (2018) 293,
arXiv:hep-ex/1712.06386.

\bibitem{ATLAS2}
G. Aad et al. (ATLAS), Eur. Phys. J. C \textbf{81} (2021) 332,
arXiv:2009.14791[hep-ex],


\bibitem{Djouadi:2005gi}
A.~Djouadi, Phys. Rept. \textbf{457} (2008) 1, arXiv:hep-ph/0503172.

\bibitem{handbook}
Report of the LHC Higgs Cross Section Working Group, S. Dittmaier et
al., Eds.,arXiv:1101.0593 [hep-ph].




\bibitem{kane}
G. L. Kane, W. W. Repko, W. B. Rolnick, Phys. Lett. B \textbf{148}
(1984) 367.

\bibitem{denys}
D. Denysiuk, PhD Thesis 2017,
https://tel.archives-ouvertes.fr/tel-01681802v2.

\bibitem{spectral}
D. Dudal, O. Oliveira, M. Roelfs, P. Silva, Nucl. Phys. B {\bf 952},
114912 (2020).

\bibitem{hill}
A. Hill and J.J. van der Bij, Phys. Rev. D\textbf{36} (1987) 3463.

\bibitem{rho1}
D.~A. Ross and M.~J.~G. Veltman, Nucl. Phys. B {\bf 95},  135
(1975).

\bibitem{rho2}
M.~J.~G. Veltman, Nucl. Phys. B {\bf 123},  89  (1977).



\bibitem{jochumtini}
J. J. van der Bij and M. Veltman, Nucl. Phys. B \textbf{231} (1984)
205.

\bibitem{nutev}
W. Bentz, I.C. Clo\"et, J.T. Londergan, A.W. Thomas, Phys. Lett. B
{\bf 693} (2010) 462.

\bibitem{partial}
P. Coloma, P. B. Denton, M.C. Gonzalez-Garcia, M. Maltoni,  T.
Schwetzg, JHEP {\bf 04}, 116 (2017).


\bibitem{fiore1992}
V. Branchina, M. Consoli, R. Fiore and D. Zappal\`a, Phys. Rev.
D{\bf 46} (1992) 75.

\bibitem{hioki1}
M. Consoli and Z. Hioki, Mod. Phys. Lett. A{\bf 10} (1995) 845.

\bibitem{hioki2}
M. Consoli and Z. Hioki, Mod. Phys. Lett. A{\bf 10} (1995) 2245.

\bibitem{erler}
J. Erler, arXiv:hep-ph/0310202, 16/10/2003.

\bibitem{wilkinson}
G. Wilkinson, arXiv:hep-ex/0205103, 30/05/2002.


\bibitem{wynhoff}
S. Wynhoff, arXiv:hep-ex/0101016, 12/01/2001.

\bibitem{schmitt}
M. Schmitt, arXiv:hep-ex/0401034v2, 25/01/2004.

\end{thebibliography}

\end{document}